\documentclass[debug,overfull]{epl}
\usepackage{amssymb}
\usepackage{amsmath}
\title{Asymmetric flux pinning in laterally nanostructured ferromagnetic /
superconducting bilayers} \shorttitle{Asymmetric flux pinning in
etc.}
\author{M.~Lange\inst{1} \thanks{E-Mail: \email{Martin.Lange@fys.kuleuven.ac.be}}
\and M.~J.~Van Bael\inst{1} \and L.~Van Look\inst{1} \and
K.~Temst\inst{1} \and J.~Swerts\inst{1} \and
G.~G\"untherodt\inst{2} \and V.~V.~Moshchalkov\inst{1} \and
Y.~Bruynseraede\inst{1}} \shortauthor{M. Lange \etal} \institute{
  \inst{1} Laboratorium voor Vaste-Stoffysica en Magnetisme, K.U.Leuven
  - Celestijnenlaan 200D, B-3001 Leuven, Belgium\\
  \inst{2} 2. Physikalisches Institut, RWTH Aachen - D-52056 Aachen, Germany
}

\pacs{74.60.Ge}{Flux pinning, flux creep, and flux-line lattice
dynamics} \pacs{74.80.Dm}{Superconducting layer structures:
superlattices, heterojunctions, and multilayers}
\pacs{75.70.-i}{Magnetic films and multilayers}

\begin{document}

\maketitle

\begin{abstract}
We investigated the pinning of flux lines in a superconducting
film by a regular array of {\em magnetic antidots}. The sample
consists of a Co/Pt multilayer with perpendicular magnetic
anisotropy in which a regular pattern of submicron holes is
introduced and which is covered by a type-II superconducting Pb
film. The resulting ferromagnetic/superconducting heterostructure
shows a pronounced asymmetric magnetization curve with respect to
the field polarity. This asymmetry clearly demonstrates that the
magnetic contribution dominates the pinning potential imposed by
the magnetic antidots on the superconducting film.
\end{abstract}

\begin{figure}
\twofigures[scale=0.47]{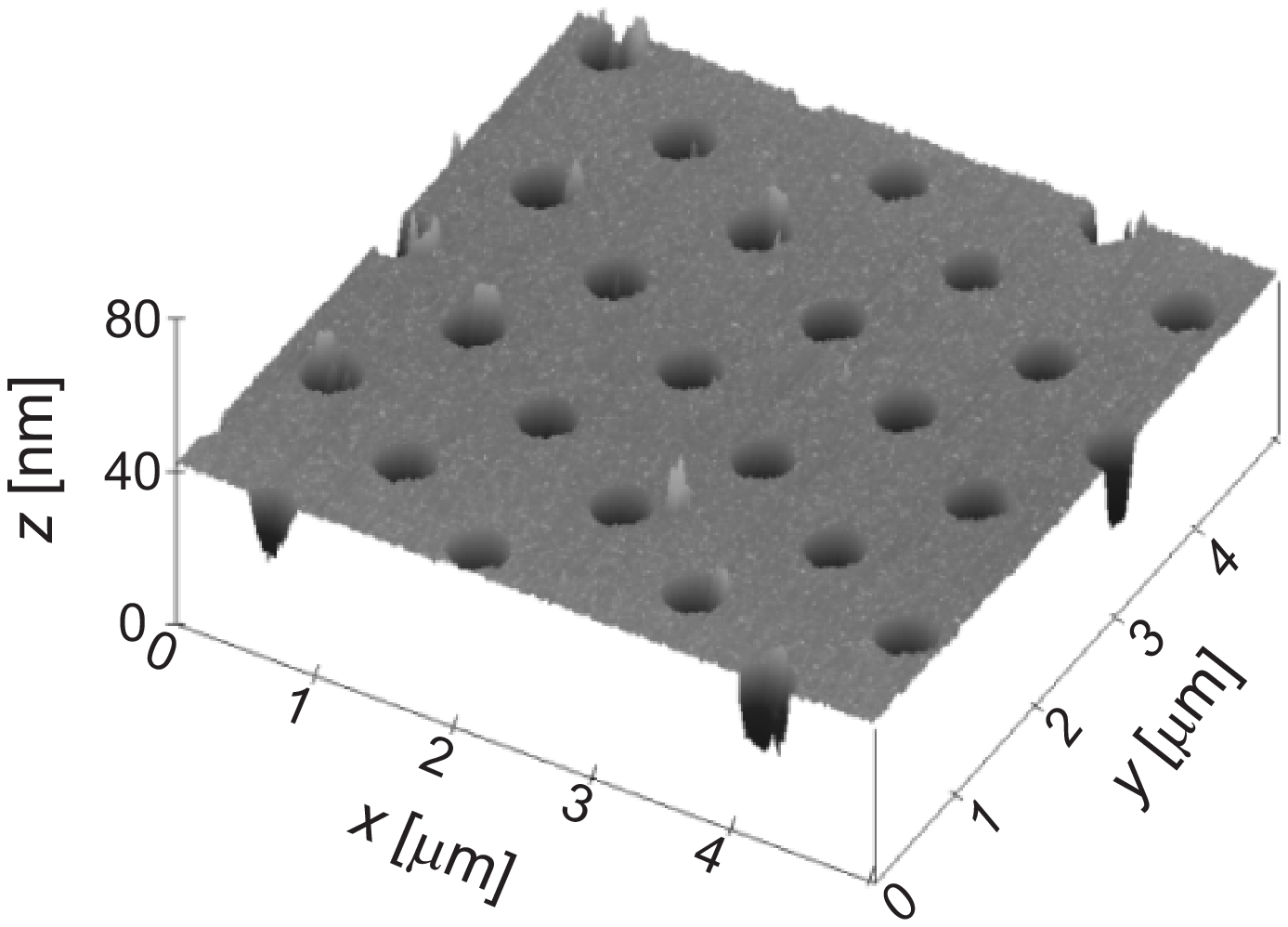}{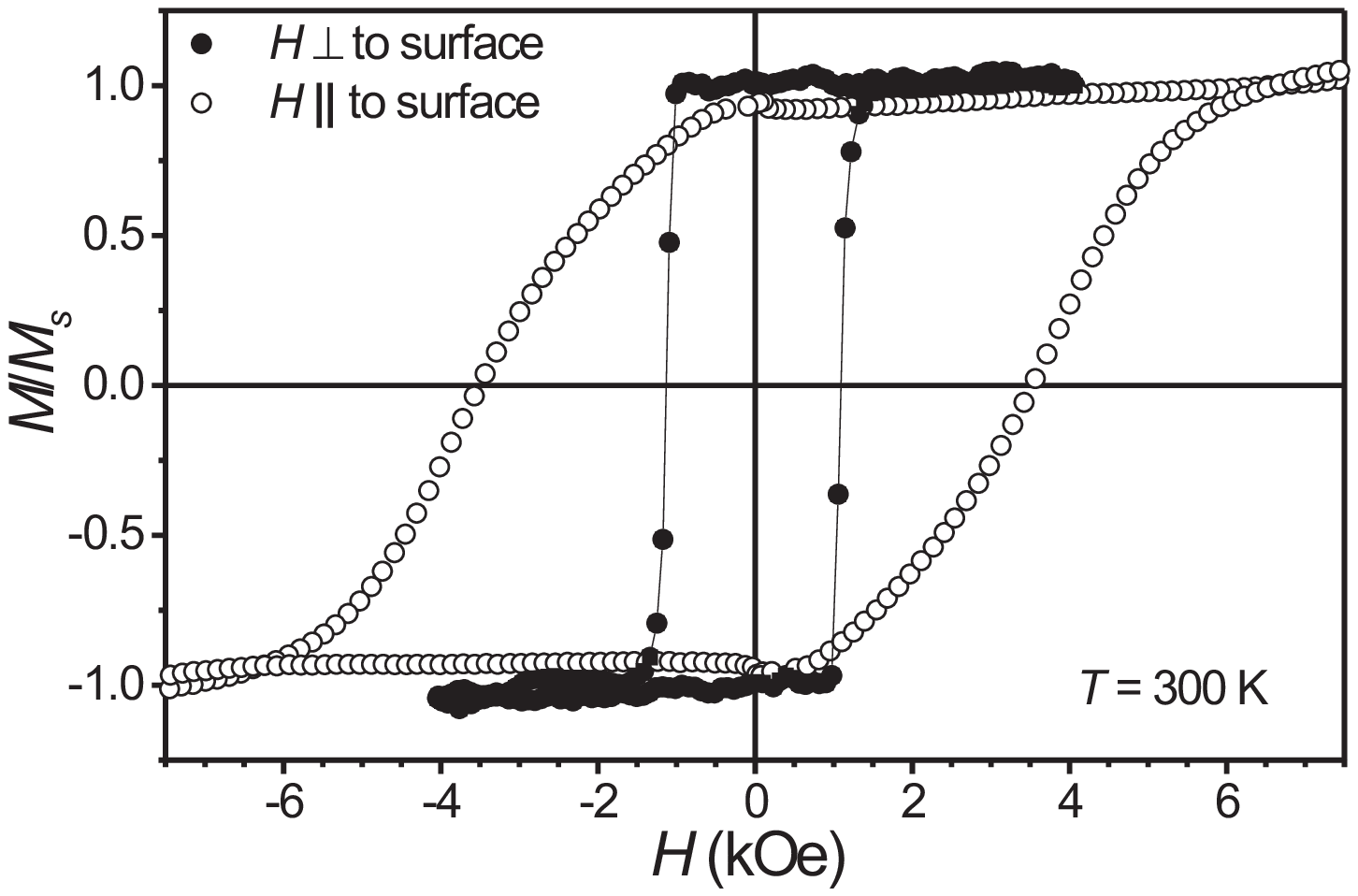} \caption{AFM image
(5~$\mu$m~$\times$~5~$\mu$m) of the Co/Pt multilayer with a square
array of antidots. The antidot array has a period of 1~$\mu$m and
the side length of an antidot is about 370~nm.} \label{afm}
\caption{MOKE hysteresis loops measured at room temperature of the
Co/Pt multilayer with antidots obtained with the field H applied
parallel ($\circ$) and perpendicular ($\bullet$) to the surface.}
\label{moke}
\end{figure}

Strong pinning of flux lines (FLs) is a prerequisite for a
superconductor to achieve a high critical current density $j_{c}$,
a condition that for instance can be realized by artificially
introducing regularly distributed pinning centres ({\em e.g.}\
antidots) in a superconducting film. The resulting
commensurability between the flux line lattice (FLL) and the
periodic pinning potential created by the regular array of
antidots in the superconductor is clearly observed as peaks or
shoulders in the curves of the magnetization $M(H)$ or $j_{c}(H)$
\cite{hebard,baertEL,baertPRL,VVM96,VVM98}, with $H$ the
perpendicularly applied magnetic field.\\ Recently, the interest
shifted to the pinning behaviour of submicron {\em magnetic}
pinning arrays in contact with a superconductor
\cite{martin97,jaccard,martin99,vanbael99,margriet,morgan,
vanbael00,lyu,sasik}. In this letter we present an experimental
study of the flux pinning in a superconducting Pb film in contact
with a Co/Pt multilayer with an array of submicron holes ({\em
magnetic antidots}). The correct choice of preparation conditions
and film thicknesses produces a Co/Pt multilayer with
perpendicular magnetic anisotropy \cite{zeper}. Several terms can
contribute to the pinning mechanism in this hybrid
ferromagnetic/superconducting system, {\em e.g.}, the corrugated
surface of the Pb film due to the deposition on top of the Co/Pt
antidot array, the high magnetic permeability of the ferromagnet
\cite{martin97,jaccard}, the direction and magnitude of the
magnetic moment \cite{morgan,vanbael00}, and the local stray field
of the ferromagnet \cite{vanbael00,vanbael99,margriet}. We show
that the $M(H)$ curve of the above described heterostructure is
strongly asymmetric, which excludes the corrugated surface of the
Pb film as the main origin of the observed matching effects.
Hence, the magnetic interaction between FLs  and pinning centres
dominates the pinning mechanism. The asymmetric $M(H)$ curves are
explained by a model based on the presence of vortices created by
the stray field of the magnetic antidots.\\ We will first describe
the sample preparation and the magnetic characterization of the
Co/Pt antidot film by means of magneto-optical Kerr effect (MOKE)
hysteresis loop and magnetic force microscopy (MFM) measurements.
In the second part, we analyse the magnetization data obtained by
a SQUID magnetometer.\\ The square array of magnetic antidots is
prepared by evaporating a $17\un{nm}$ thick Co/Pt multilayer in a
resist mask on an amorphous \chem{SiO_2} substrate held at room
temperature. The resist mask is prepatterned by electron-beam
lithography, and after the deposition of the Co/Pt multilayer the
resist is removed using a standard lift-off procedure. The
patterned area is about $9\un{mm^{2}}$. The Co/Pt multilayer is
deposited in an MBE system by e-beam evaporation with typical Pt
and Co deposition rates of about $0.06\un{nm/s}$, controlled by a
quartz crystal oscillator. The multilayer consists of a
$2.8\un{nm}$ Pt base layer to improve the perpendicular anisotropy
\cite{Li}, and a multilayer structure of
[Co($0.4\un{nm}$)/Pt($1.0\un{nm}$)]$_{10}$.\\ An atomic force
microscopy (AFM) measurement of the antidot array is shown in
fig.~\ref{afm} and reveals a well-defined square arrangement of
the antidot array with a period of $1\un{\mu m}$.
The antidots have a square shape with rounded corners and a side
length of about $370\un{nm}$.\\ The Co/Pt antidot array is covered
with a continuous type-II superconducting Pb film to study the
flux pinning properties of the magnetic antidots. In order to
prevent the direct influence of the proximity effects between Pb
and Co/Pt, a $10\un{nm}$ insulating amorphous Ge layer is
deposited first with a growth rate of $0.2\un{nm/s}$, then the
$50\un{nm}$ Pb film is evaporated at a substrate temperature of
$77\un{K}$ with a growth rate of $1.0\un{nm/s}$ and finally, the
sample is covered with a $30\un{nm}$ Ge layer for protection
against oxidation. Note that the Pb layer is significantly thicker
than the Co/Pt multilayer, which ensures a complete coverage of
the magnetic antidots.\\ To confirm that the easy axis of
magnetization of the Co/Pt multilayer with antidots is
perpendicular to the surface, we measured MOKE hysteresis loops at
room temperature in two different field configurations before the
patterned multilayer was covered with the Ge/Pb/Ge trilayer (see
fig.~\ref{moke}). With the field $H$ applied perpendicular to the
sample plane, the loop has a rectangular shape with a saturation
field of $H_{s}=1.5\un{kOe}$ and a coercive field of
$H_{c}=1.1\un{kOe}$. When $H$ is applied parallel to the surface,
we obtain $H_{c} = 3.5\un{kOe}$ and $H_{s} \approx 7\un{kOe}$.
These results indicate that the easy axis of magnetization is
perpendicular to the surface. At low temperatures ($T=5\un{K}$),
SQUID measurements on a Co/Pt reference multilayer without any
patterning show that the 100~\% remanent magnetization of the
hysteresis loop obtained with $H$ perpendicular to the surface is
preserved, while $H_{c}$ is slightly enlarged.\\ For the
interpretation of the flux pinning phenomena in the
ferromagnetic/superconducting hybrid structure, it will be
important to obtain microscopic information about the domain
structure and the related stray field patterns of the Co/Pt
antidot array. For that reason we have carried out MFM
measurements of the patterned multilayer without Ge/Pb/Ge trilayer
using a Digital Instruments Nanoscope~III system at room
temperature and zero field.
\begin{figure}
\onefigure[scale=0.99]{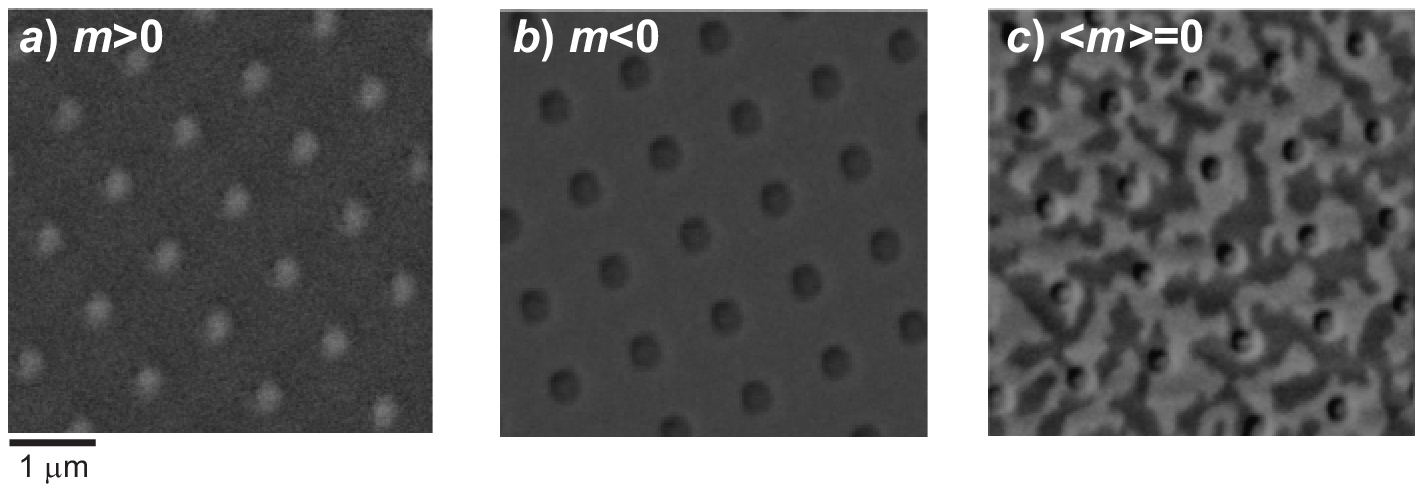} \caption{MFM images ($5\un{\mu m}
\times 5\un{\mu m}$) measured in $H=0$ of the Co/Pt multilayer
with antidots {\em a})~after magnetization in $H = +10\un{kOe}$,
{\em b})~after magnetization in $H = -10\un{kOe}$ and {\em
c})~after demagnetization.} \label{mfm}
\end{figure}
The magnetic moment of the tip is pointing perpendicular to the
surface, which makes it sensitive to the perpendicular component
of the stray field emanating from the surface. The experiments are
carried out using the tapping/lift$^{\textrm{TM}}$ mode
\cite{zhong} with a typical scan height of $50-80\un{nm}$ above
the sample surface. Figure~\ref{mfm} shows three $5\un{\mu m}
\times 5\un{\mu m}$ MFM scans of the sample in three different
magnetic states. The magnetic moments ${\bm m}$ of the Co/Pt
multilayer are aligned into different directions before the MFM
measurement by applying a field perpendicular to the surface of
(fig.~\ref{mfm}{\em a})~$H = 10\un{kOe}$ ($m=|{\bm m}|>0$), and
(fig.~\ref{mfm}{\em b})~$H = -10\un{kOe}$ ($m<0$). In
fig.~\ref{mfm}{\em c}, the sample was demagnetized in an
out-of-plane field oscillating around zero with decreasing
amplitude before the MFM measurement ($\langle m \rangle=0$). In
fig.~\ref{mfm}{\em a}, bright spots are observed at the position
of the antidots, which can be explained by the fact that the
out-of-plane components of the stray field ${\bm h}$ above the
antidots and above the multilayer are oriented in a mutually
opposite direction. When $m<0$ (fig.~\ref{mfm}{\em b}), the spots
appear in a darker colour, corresponding to the other polarity of
${\bm h}$ compared to the $m>0$ state. In fig.~\ref{mfm}{\em c},
the bright and dark regions between the antidots can be identified
as magnetic domains in the Co/Pt multilayer, where the
magnetization is directed either parallel or antiparallel to the
tip magnetization. The observed domain structure is typical for
thin magnetic films with perpendicular anisotropy (see {\em e.g.}\
\cite{allenspach}) and the average domain size amounts to a few
$100\un{nm}$. The contrast at the edges of the antidots might be
due to some tip effects because of the topography. It is clear
that the created stray field pattern in fig.~\ref{mfm}{\em c} is
different from the one that is found when the sample is magnetized
(fig.~\ref{mfm}{\em a} and \ref{mfm}{\em b}).\\ We will now
investigate how the different magnetic states of the sample
influence the flux pinning phenomena in the superconducting Pb
layer. After covering the magnetic antidots with the type-II
superconducting Pb film, $M(H)$ curves have been measured below
the critical temperature $T_{c}$ of the Pb film in a Quantum
Design SQUID magnetometer with $H$ perpendicular to the film
plane. From $M(T)$ measurements we determined $T_{c}=7.20\un{K}$.
Figure~\ref{mh} shows the upper branches ($M>0$) of the $M(H)$
loops of the sample at $T = 7\un{K}$. The lower branches ($M<0$)
are mirror images of the upper branches with respect to the field
axis.
\begin{figure}
\onefigure[scale=0.66]{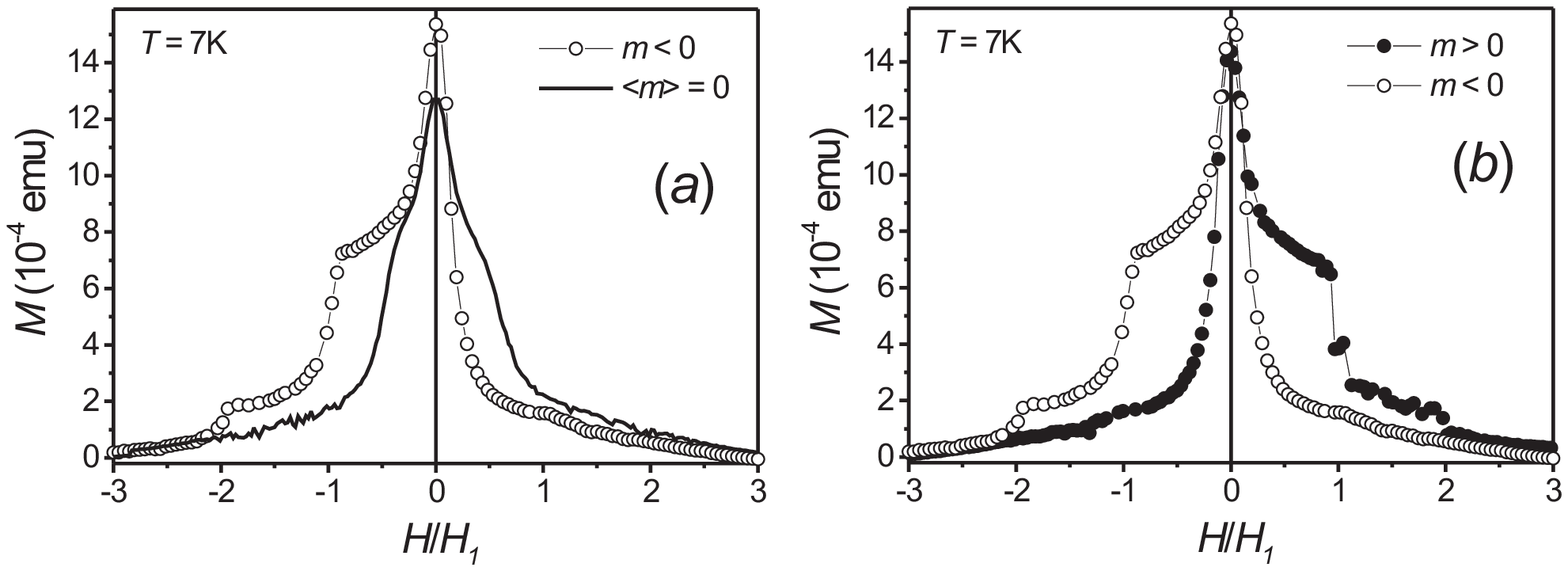} \caption{SQUID magnetization
curves at $T=7\un{K}$ of a superconducting Pb film on top of a
magnetic Co/Pt antidot lattice, with the antidot lattice in
different magnetic states: {\em a})~($\circ$)~$m<0$ and full line
$\langle m \rangle=0$, {\em b})~($\circ$)~$m<0$ and
($\bullet$)~$m>0$.} \label{mh}
\end{figure}
The field scale is normalized to the first matching field $H_{1} =
\phi_{0}/(1 \un{\mu m})^{2} = 20.67\un{Oe}$, at which $H$
generates exactly one flux quantum $\phi_{0}$ per unit cell of the
magnetic antidot array \cite{baertEL,baertPRL,VVM96,VVM98}. $H$ is
always much smaller than the coercive field $H_{c}$ of the Co/Pt
multilayer with antidots, so that the magnetic state of the
multilayer remains unchanged for each of the displayed curves in
fig.~\ref{mh}. The ferromagnetic contribution of the Co/Pt
multilayer to the total magnetization $M$ results in an offset of
the order of $\sim 10^{-5}\un{emu}$ depending on its magnetic
state. The preservation of the magnetic state is confirmed by an
unchanged offset after measuring the $M(H)$ curves.\\ It is clear
from fig.~\ref{mh}{\em a} that the $M(H)$ curve for $m<0$ is {\em
strongly asymmetric} with respect to the polarity of the applied
field, very similar to what was observed recently for a
superconducting film on top of a lattice of magnetic dots with
out-of-plane magnetization \cite{morgan,vanbael00}. For the $m<0$
state, the magnetization $M$ has a larger value at negative fields
than at the corresponding positive fields, indicating a much
stronger flux pinning when $H$ and $m<0$ are pointing in the same
direction. The field polarity dependent pinning strength can also
be noticed in the matching effects, which appear at integer
negative matching fields at $H/H_{1}=-1$ and $-2$, while for their
positive counterparts, only a small deviation from the smooth
curve is visible at $H/H_{1}=1$. No matching effect can be seen at
$H/H_{1}=2$. In the $\langle m \rangle=0$ state, a symmetric
$M(H)$ curve is obtained (solid line in fig.~\ref{mh}{\em a}),
with only weak matching effects appearing as shoulders at $H/H_{1}
= +1/2$ and $-1/2$. Figure~\ref{mh}{\em b} illustrates that the
asymmetry of the $M(H)$ curves is {\em reversed} when $m>0$. The
enhanced matching effects and the larger value of $M$ now appear
at positive fields.\\ Figure~\ref{tempde} shows the temperature
dependence of the $M(H)$ curves with the magnetic antidots in the
$m>0$ state.
\begin{figure}
\twofigures[scale=0.51]{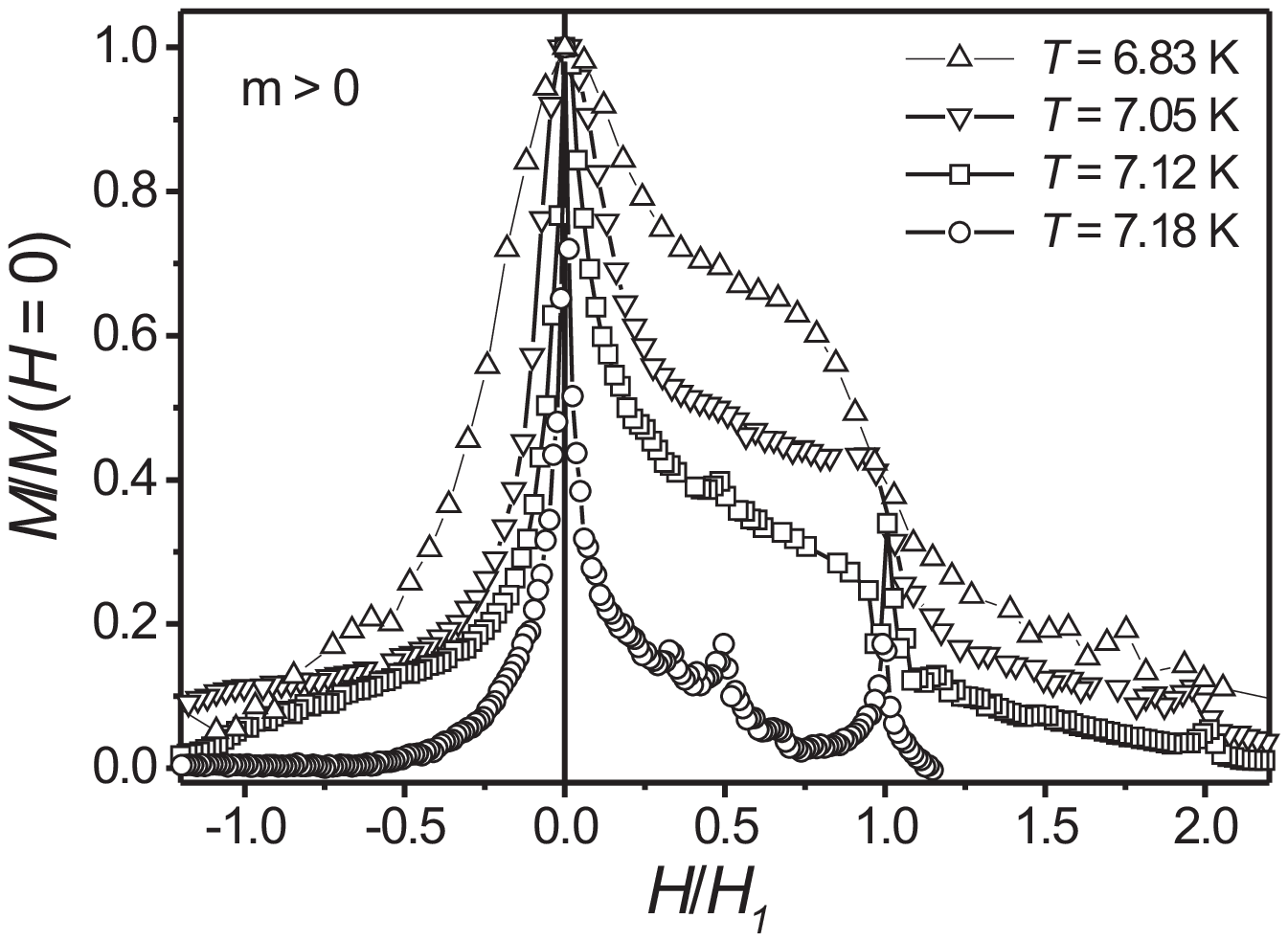}{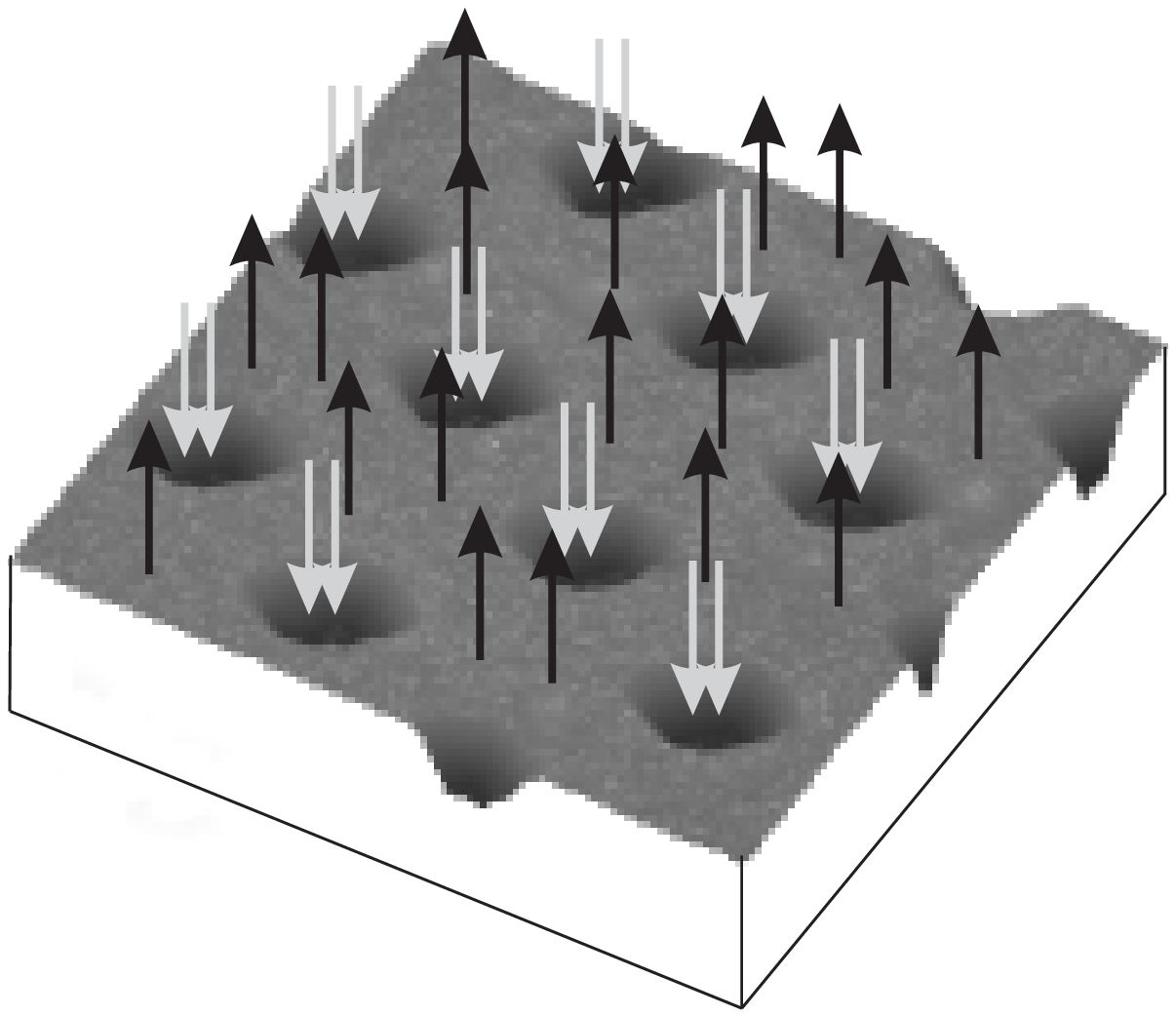} \caption{Magnetization
curves of a superconducting Pb film on top of a Co/Pt antidot
lattice ($m>0$) at different temperatures:
($\triangle$)~$T=6.83\un{K}$, ($\nabla$)~$T=7.05\un{K}$,
($\square$)~$T=7.12\un{K}$ and ($\circ$)~$T=7.18\un{K}$.}
\label{tempde} \caption{Schematic drawing to illustrate the
proposed model for $m>0$ and $H=0$: vortices (black arrows) and
antivortices (grey arrows) are induced in the superconductor by
the stray field of the patterned Co/Pt multilayer. The value of
the stray field corresponds to $2 \phi_{0}$ per unit cell of the
antidot array. The antivortices act as a regular array of pinning
centers for external flux lines.} \label{model}
\end{figure}
At temperatures close to $T_{c}$, we observe additional rational
matching effects at positive fields ($H/H_{1} = 1/3, 1/2$ and
$2/3$ at $T=7.18\un{K}$, and $H/H_{1} = 1/2$ at $T=7.12\un{K}$),
no matching effects can be seen at negative fields. If the
temperature is decreased, the rational matching effects disappear
and the integer matching effects are smeared out ($T=6.83\un{K}$).
This is also observed in thin superconducting films with lattices
of antidots \cite{baertEL}. However, the $M(H)$ curve keeps its
asymmetry down to the lowest measured temperature of $5\un{K}$.\\
The $M(H)$ data show that vortices are much stronger pinned when
$H$ and $m$ have the same polarity. This observation raises the
question: which contributions to the pinning potential are
responsible for the asymmetry ?\\ The {\em surface modulation} of
the Pb film cannot explain the asymmetry of the $M(H)$ curves,
since its contribution to the pinning potential is not depending
on the direction of $H$. For the same reason, the high {\em
magnetic permeability} of the ferromagnetic material can not be
the origin of the asymmetry. For magnetic dots with perpendicular
anisotropy as pinning centres, the asymmetric behaviour has been
related to the interaction between the magnetic moments ${\bm m}$
of the magnetic dots and the applied field ${\bm H}$, in terms of
a {\em magnetic interaction energy} $E = - {\bm m} \cdot {\bm H}$
\cite{morgan,vanbael00}. Evaluating this term for the magnetic
antidots, the clearest matching effects should appear when $H$ and
$m$ have opposite polarity (in this case FLs are forced into the
position of the antidots, which are acting as sharply defined
pinning centers). This is in contrast to what is observed in this
experiment, therefore the term $E = - {\bm m} \cdot {\bm H}$
cannot dominate the pinning potential.\\ Recent vortex imaging
experiments investigating in-plane magnetized dots in contact with
a superconductor have shown that the stray field of the dots can
induce vortex-like structures in the superconductor. These induced
vortices play a crucial role in the flux pinning properties of the
dots \cite{margriet}. Also for magnetic dots with out-of-plane
magnetization, the existence of induced vortices has been
predicted by theory \cite{marmorkos,lyu,sasik}. Magnetic antidots
have the inverse geometry of magnetic dots, therefore the stray
fields of dots and antidots are very similar to each other. We
will now show that the asymmetric $M(H)$ curves in fig.~\ref{mh}
can be explained by the presence of such stray field induced
vortices in the superconductor. \\ Based on the above
considerations for magnetic dots, we propose the following model
when magnetic antidots are used as pinning centers: The presence
of the stray field will create supercurrents in the
superconductor. In order to estimate the magnitude of the stray
field, we have carried out magnetostatical calculations (based on
the theory in \cite{jackson}) for magnetic antidots with the same
geometry and magnetization as in this experiment. At a height of
$35\un{nm}$ above the surface of the Co/Pt multilayer, the amount
of flux that is created by the ferromagnet corresponds to about $2
\times \phi_{0}$ ($\phi_{0}$ is the flux quantum) in each unit
cell of the magnetic antidot array. Hence, it appears reasonable
to assume that the local stray fields cannot be completely
shielded by the Meissner currents and at least one vortex per unit
cell of the magnetic antidot array is created by the stray field.
It is possible that different unit cells contain different amounts
of flux quanta, {\em e.g.}\ when the average created flux in one
unit cell above the ferromagnet is $r \times \phi_{0}$, with $r$ a
non-integer number.\\ In the $m>0$ state, we assume that these
induced vortices with a local magnetic field $h>0$ will be located
at the interstitial positions of the antidot array (see black
arrows in fig.~\ref{model}). Since stray field emanating from the
Co/Pt multilayer must return somewhere, antivortices with $h<0$
will appear at the position of the antidots (grey arrows in
fig.~\ref{model}). The existence of vortex-antivortex pairs
induced by stray field is also predicted in the case of magnetic
dots with out-of-plane magnetization \cite{lyu,sasik}. Vortices
and antivortices can not annihilate, since the vortices must be
located at the interstitial positions, whereas the antivortices
are fixed at the antidots. The number of antivortices and vortices
in the superconductor in zero field must be the same (neglecting
effects at the sample boundary). \\ We will now discuss how the
behaviour of the system in an applied magnetic field can be
described by means of the above model. To avoid any confusion
concerning the use of the terms {\em vortex, antivortex} and {\em
FL}, we will always use the terms {\em vortices} and {\em
antivortices} when they are induced by the magnetic Co/Pt
multilayer (vortices located at the interstices and antivortices
at the antidots) and {\em FLs} when they are created by an
externally applied field. When $H \neq 0$ is applied, the
additional FLs will interact with the vortices and antivortices.
If FLs and vortices have the same field polarity, they have a
repulsive interaction, while FLs and vortices with opposite field
polarity attract each other. The antivortices are fixed at the
position of the antidots and therefore form a strongly regular
pinning potential for the FLL, with a period corresponding to the
period of the antidot array. The vortices at the interstitial
positions are more mobile and do not form such a regular array,
see fig.~\ref{model}.\\ When $H>0$ and $m>0$, FLs and antivortices
have an attractive interaction and therefore, matching effects and
an enhanced value of $M$ are observed. When $H<0$ and $m>0$, FLs
and antivortices have the same polarity and repel each other, so
that the FLs are forced into the interstitial positions. Hence, a
lower value of $M$ and only weak matching effects are observed
compared to the case when $H>0$ and $m>0$. This model also
explains the reversal of the asymmetry when the film is magnetized
in the opposite direction ($m<0$). The almost symmetric $M(H)$
curve obtained for demagnetized magnetic antidots ($\langle m
\rangle=0$) can be explained by the absence of an asymmetric
pinning potential. Due to the domain structure (see
fig.~\ref{mfm}{\em c})), antivortices can also appear at the
interstitial positions. However, the weak matching effects at
$H/H_{1}=+1/2$ and $-1/2$ indicate that the pinning potential
still has a certain periodicity, {\em e.g.}, due to non-magnetic
contributions such as the surface modulation of the Pb.\\ In
conclusion, a lattice of antidots in a magnetic film with
perpendicular anisotropy creates a strong pinning potential for
flux lines in a type-II superconductor. Pronounced asymmetric
$M(H)$ curves are observed, indicating significantly enhanced
pinning when $m$ and $H$ are parallel. To explain this asymmetry,
a qualitative model is proposed which is based on the presence of
vortices and antivortices induced by the stray field of the
magnetic antidots. A quantitative theoretical treatment and
further experiments by local vortex imaging techniques are needed
to confirm this model.

\acknowledgments The authors would like to thank R. Jonckheere for
the preparation of the resist pattern and U. May, P. Miltenyi and
J. Keller for the assistance with the MBE. This work was supported
by the Belgian Inter-University Attraction Poles (IUAP) and
Flemish Concerted Research Actions (GOA) programs, by the ESF
"VORTEX" program and by the Fund for Scientific Research-Flanders
(FWO). MJVB and KT are Postdoctoral Research Fellows of the FWO.

\newpage
\section{erratum}
\begin{figure}
\onefigure[scale=0.68]{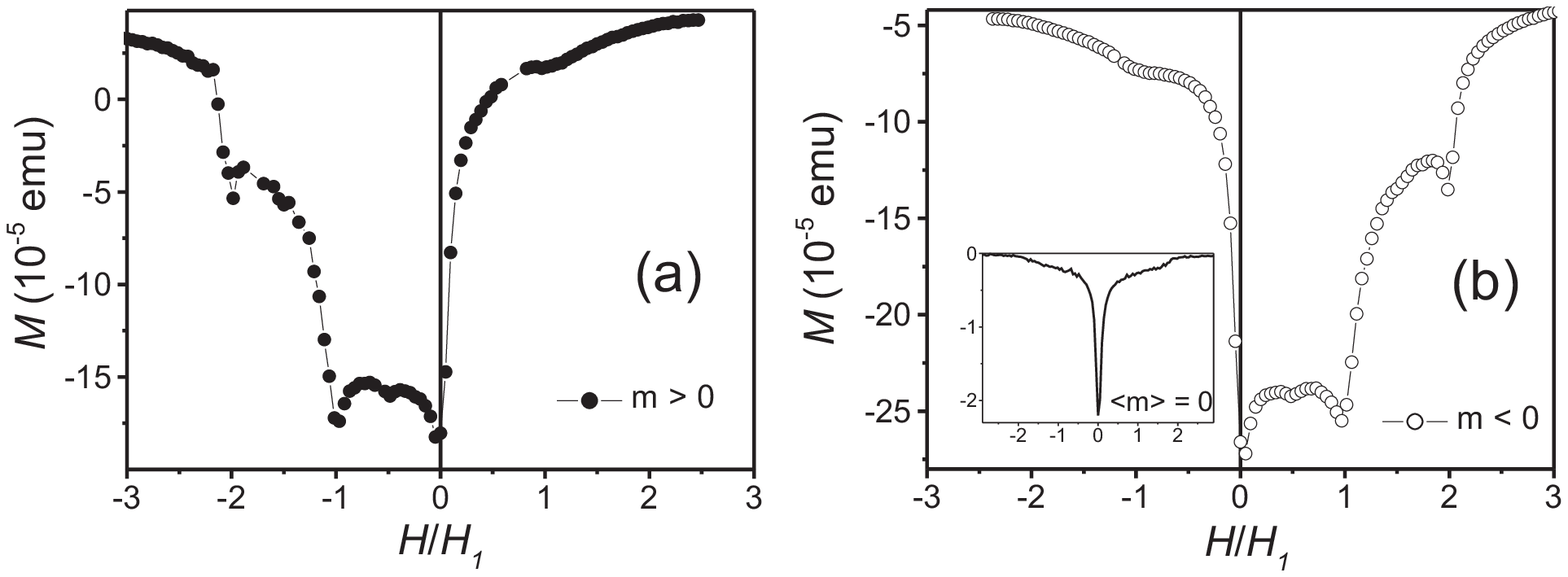} \caption{SQUID magnetization
curves at $T=7.05$~K of a superconducting Pb film on top of a
magnetic Co/Pt antidot lattice, with the antidot lattice in
different magnetic states: {\em a})~($\bullet$)~after saturation
in a positive field ($m>0$); {\em b})~($\circ$)~after saturation
in a negative field ($m<0$); and in the inset of {\em b})~after
demagnetization ($\langle m \rangle =0$).} \label{newmh}
\end{figure}

We have found recently a technical mistake in presenting the
magnetization curves of the nanostructured
ferromagnetic/superconducting bilayer in figs.~4 and 5 of our
letter \cite{lange}. As a result, the discussion of these figures
and of the pinning potential on p.~650-651 should be changed. The
right magnetization curves for the three magnetic states of the
sample are shown in figure~\ref{newmh}.
The enhanced matching effects are observed when $H$ and $m$ have
mutual {\em opposite} polarity and the matching effects are
suppressed when $H$ and $m$ have the same polarity, contrary to
the model that was used in Ref.~\citen{lange} to explain the
results and that was illustrated by fig.~6. Consequently the model
and fig.~6 are not relevant anymore for the description of the
pinning phenomena in the nanostructured
ferromagnetic/superconducting bilayer studied in
Ref.~\citen{lange}.\\ Consistent with the main conclusions of
Ref.~\citen{lange}, the strong asymmetry should unambiguously be
attributed to the magnetic contributions to the pinning potential
and can be related to the magnetic interaction energy $E = - {\bm
m} \cdot {\bm H}$ \cite{morgan} or an interaction between flux
lines and stray field-induced supercurrents around the antidots
\cite{vanbael}. A more detailed discussion of these results will
be reported elsewhere.

\end{document}